# New Buckled Honeycomb Lattice Compound $Sr_3CaOs_2O_9$ Exhibiting Antiferromagnetism above Room Temperature

Gohil S. Thakur[a,b], T. C. Hansen[c], Walter Schnelle[a], Shuping Guo[d], Oleg Janson[d], Jeroen van den Brink[d], Claudia Felser[a], and Martin Jansen[a,e] *

[a] Max Planck Institute for Chemical Physics of Solids, Nöthnitzer Strasse 40, 01187 Dresden, Germany

[b] Faculty of Chemistry and Food Chemistry, Technical University, 01069 Dresden, Germany

[c] Institut Laue-Langevin, 71 avenue des Martyrs, 38000 Grenoble, France

[d] Institute for Theoretical Solid State Physics, Leibniz IFW Dresden, 01069 Dresden, Germany

[e] Max Planck Institute for Solid State Research, Heisenberg Strasse 1, 70569 Stuttgart, Germany

**ABSTRACT:** Synthesis, crystal structure and magnetic properties of a new 2:1 ordered triple perovskite $Sr_3CaOs_2O_9$ are reported. The compound crystallizes in $P2_1/c$ space group and features a unique buckled honeycomb lattice of osmium. It exhibits long-range antiferromagnetic ordering with a high Néel temperature of 385 K as confirmed by susceptibility, heat capacity and neutron diffraction measurements and is electrically insulating. This compound is also the first example of a 2:1 ordered osmate perovskite. Theoretical investigations indicate that $Sr_3CaOs_2O_9$ features a sizeable antiferromagnetic exchange between the puckered planes resulting in a high $T_N$. The magnetic properties of the known compound $Sr_3CaRu_2O_9$ are elucidated in comparison. It shows antiferromagnetic order below $T_N \approx 200$ K.

## 1. INTRODUCTION

Antiferromagnetic insulating materials offer exciting prospects for device industry as active components in spin valve and opto-spintronic devices.[1–5] In such implementations the spin of the electrons is used, rather than the charge, to store information, thereby circumventing the problem of Joule heating associated with the flow of electrons. Also, by virtue of the zero net moment displayed by antiferromagnets, electron spins produce no stray field and are practically insensitive to external magnetic disturbances which allows for a much denser stacking of components.[4,5] Antiferromagnetic insulators are also expected to have terahertz writing speed leading to ultrafast operations.[6] Air stable room-temperature antiferromagnetic insulators are thus highly desirable. Perovskite oxides present an excellent avenue to explore such materials due to their ubiquitous nature, excellent stability, structural robustness, and wide range of atomic flexibility. In ordered double ($A_2BB´O_6$) or triple ($A_3BB´_2O_9$) perovskites, the B sites appear to be very versatile as a variety of metal atoms ranging from alkaline earth (Mg, Ca, Sr, Ba), alkali (Li, Na, K) and transition metals can be substituted or doped at those sites, leading to a wide range of magnetic and electrical properties.[7,8] Same stands true for the A-sites although to a limited extent. The resulting compositional and structural degrees of freedom enable to realize various 1:1 or 2:1 ordered perovskites, given that certain size and charge requirements are fulfilled. Here we report on a 2:1 ordered semiconducting triple perovskite $Sr_3CaOs_2O_9$ exhibiting a high Néel temperature. This compound is also exceptional in another respect. Conventional perovskites with 2:1 ordering with $d^0$ (nonmagnetic) metals (Nb, Ta, W etc.) are abundant while the ones with partially filled $d$ orbitals (magnetically active) are quite uncommon.[9,10] $Sr_3CaRu_2O_9$ has been the first reported example of a 2:1 ordered conventional perovskite containing a non $d^0$ ion at the B site,[11] and $Sr_3CaIr_2O_9$ being the only other isostructural compound reported till date.[12] In contrast, there are many compounds with this composition known with Ba occupying the A site. Accordingly, they all derive from the hexagonal perovskite structure type featuring face sharing $M_2O_9$ dimers (M = Ru and Os), e.g. $Ba_3MOs_2O_9$ (M = □, Li, Na, Cu and Zn) and $Ba_3MRu_2O_9$ (M = Li, Na, Cu and rare-earth).[13–21]

In continuation to our expedition to find compounds containing osmium and featuring magnetic ordering transitions at high temperature,[22–25] we report the new compound $Sr_3CaOs_2O_9$, crystallizing in a conventional monoclinic triple-perovskite structure. $Sr_3CaOs_2O_9$ orders antiferromagnetically at $T_N \approx 385$ K. The previously known isostructural compound $Sr_3CaRu_2O_9$, magnetic properties of which were hitherto unreported, has also been studied using magnetization techniques. It is found to be antiferromagnetic with $T_N \approx 200$ K.

## 2. EXPERIMENTS AND PROCEDURES

**Synthesis.** Black polycrystalline powder samples of $Sr_3CaOs_2O_9$ were prepared along a standard solid-state sealed tube method. Stoichiometric amounts of highly pure SrO$_2$ (Sigma Aldrich, 98%), SrO (Sigma Aldrich, 97 %), CaO (Alfa Aesar, 99.9 %) and OsO$_2$ (Alfa Aesar, min 83 % Os) (total of 0.2 g) were ground, pelletized and kept in a corundum crucible and placed in a quartz tube, which was sealed off under vacuum. Since the commercially available OsO$_2$ may contain a large amount of Os metal, it was further oxidized prior to use by heating in presence of PbO$_2$, contained in a separate crucible, in an evacuated quartz tube at 753 K for 2 days. This treatment yielded highly pure and crystalline OsO$_2$. All the chemical manipulations were performed in an Ar-filled glove box (O$_2$ and H$_2$O <1 ppm). The tube was then heated at 1073 K for 48 hours at a heating and cooling rate of 100 K per hour in a tubular furnace placed inside a fume hood. The tubes were cracked–open in a well-ventilated fume hood. The product is stable in air for months. A total of 2 grams of sample for neutron studies was prepared in the same manner in several smaller batches. Sr$_3$CaRu$_2$O$_9$ was prepared by heating appropriate amounts of respective alkaline earth carbonates and RuO$_2$ in air at 1473 K as reported previously.[11]

**Instruments for physical property measurements.** Laboratory powder X-ray diffraction (PXRD) data were collected at room temperature on a HUBER G670 imaging plate Guinier camera with Cu-K$\alpha$1 radiation ($\lambda$ = 1.5406 Å), in a 2$\theta$ range of 5–100 degrees.

Crystal structure was refined by a Rietveld profile fit of the powder X-ray data using the TOPAS-4.2.0.2 (AXS) program[26] starting with the Sr$_3$CaRu$_2$O$_9$ structure[11] as the initial model. Refined parameters were cell constants, Wyckoff positions, scale factor, zero point of $\theta$, sample displacement (mm), background as a Chebyshev polynomial of 20th degree, 1/$x$ function, crystallite size and micro-strain. The structure was satisfactorily refined in the monoclinic $P2_1/c$ space group.

Neutron diffraction data was recorded on the high-intensity powder diffractometer D20 at the HFR of ILL in Grenoble (France) in the frame of the experiment EASY-686.[27] High resolution patterns were recorded at ambient temperature using a vertically focusing germanium (hhl) monochromator at a take-off angle of 118° at 1.36 Å (117) and 1.87 Å (115). A 2$\theta$ scan (61 steps of step width 0.05° 2$\theta$, the position sensitive detector's angular pitch being 0.05° as well, 60 s counting time per step, thus 61' in total) has been used to obtain very clean patterns through a maximum-likelihood approach by redundant measurements at different 2$\theta$ angles. Medium resolution data (without detector scan) was recorded at 2.42 Å using a pyrolytic graphite (002) monochromator at 42° take-off whilst heating the sample on a linear temperature ramp from 1.8 to 395 K. Further, 2$\theta$ scans (61 steps at 0.05°, 120 s per step, thus 122' in total) at lowest temperature (1.8 K) and just above transition temperature (395 K) were recorded to provide good differential data (despite the high temperature difference) for a detailed analysis of the magnetic structure. Crystallographic structure data obtained from the neutron powder refinement (at 1.36 Å) has been deposited with Fachinformationszentrum Karlsruhe, D-76344 Eggenstein-Leopoldshafen (Germany) and can be obtained on quoting the depository number CSD-2153716.

Magnetization data of powder samples were collected in the temperature range 2-350 K in applied magnetic fields $\mu_0 H$ = 3.5 and 7.0 T on a MPMS-XL7 and from 300 to 750 K on a MPMS-3 magnetometer (both from Quantum Design). Isothermal magnetization curves were recorded for magnetic fields between ±7 T at $T$ = 2 K. Diamagnetic corrections were applied.[28]

**Computational methods**: First-principles density functional theory (DFT) calculations were carried out by using the experimental structure of Sr$_3$CaOs$_2$O$_9$ with two codes, Vienna *ab initio* Simulation Package VASP and all-electron full-potential local-orbital code FPLO.[29] Total energy calculations were performed in a plane-wave basis set using the projector augmented wave (PAW)[30] method under a gradient generalized approximation (GGA)[31] with PBE exchange-correlational functional and employing DFT+U[32] methods as implemented in the VASP. We used an energy cutoff of 500 eV and and a 2×5×4 gamma-centered Monkhorst-Pack[33] k-point mesh to sample the Brillouin zones. All structures were relaxed with all the forces and tensors are below 0.01 eV Å$^{-1}$ and 0.2 kbar, respectively. Hubbard $U$ correction and exchange interaction parameter Hund $J$ used for Os[34] were 2 and 0.4 eV, respectively. The FPLO was used for the tight-bonding Wannier function analysis. The Brillouin zone was sampled with a 3×5×7 mesh. Theoretical crystal structure and magnetic moments of Os are listed in Table S10 and S11 in the SI, respectively.

## 3. EXPERIMENTAL RESULTS AND ANALYSIS

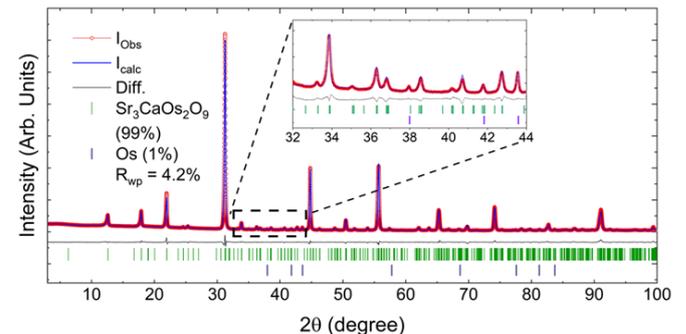

Figure 1: Laboratory PXRD data (red line) along with the Rietveld fit (blue line) of Sr$_3$CaOs$_2$O$_9$. Black line is the difference curve and vertical bars represent the Bragg reflections corresponding to respective phases. Inset shows the magnified area in the 2$\theta$ range 32-44°.

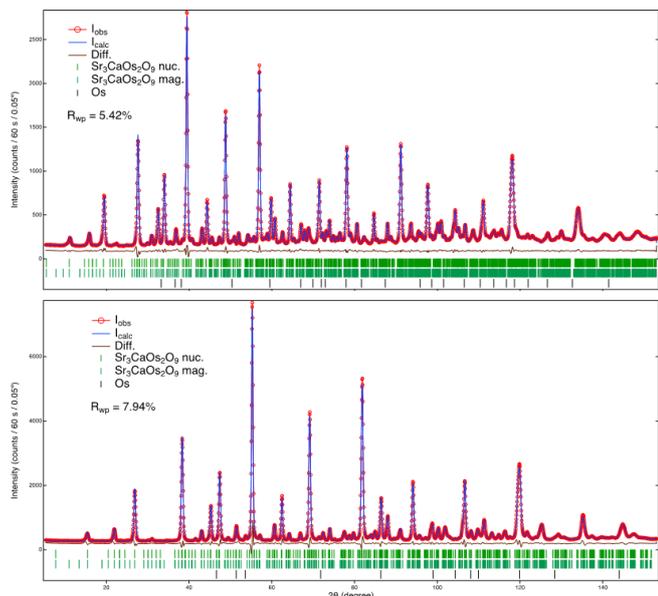

Figure 2: Room temperature powder neutron data (red circles) recorded at wavelengths 1.87 Å (upper) and 1.36 Å (lower) and their Rietveld fits (blue lines). Brown lines are the difference curves and upper and lower green vertical bars represent the Bragg reflections corresponding to respective nuclear and magnetic cells of $Sr_3CaOs_2O_9$.

3.1 **Structure.** A new 2:1 ordered osmium-containing perovskite $Sr_3CaOs_2O_9$ has been synthesized. The compound appears as black air stable microcrystalline powder. The elemental ratio of metals was confirmed by SEM-EDX analysis (see Figure S1 in supplementary information (SI)). On heating up to 1450 K in argon atmosphere, $Sr_3CaOs_2O_9$ only partially decomposes to $Sr_3CaOs_2O_9$ (~85 %) $Sr_2CaOsO_6$ (~10 %) and Os metal 5% (see Figures S2 and S3 in SI). The structure was successfully refined using room temperature powder X-ray data in the monoclinic space group $P2_1/c$ (#14) which was later confirmed by neutron diffraction. The structure of $Sr_3CaRu_2O_9$ was used as the initial model for refinement. Satisfactory $R_{wp}$ of 4.02 % was obtained after the final refinement cycle. A trace amount (<1 %) of elemental osmium is always apparent in the diffraction patterns of $Sr_3CaOs_2O_9$ and was therefore included in the refinement (Figure 1). Crystallographic details and atomic coordinates obtained after Rietveld refinement of the powder X-ray data are presented in Table S1 and S2 in the SI.

To confirm the crystal structure and site occupancies of atoms, especially the lighter ones, we employed high resolution neutron diffraction measurements at two different wave lengths 1.87 and 1.36 Å. The structure refinement was carried out on the conjointly applied room temperature data obtained at two wavelengths (Figure 2). Although, from the X-ray data, we observed a complete ordering at the B sites with full occupancies at all the sites and reasonable isotropic thermal parameters, in the neutron refinement, despite a small absorption correction, the (isotropic) thermal displacement parameter of one of the calcium positions remained unphysically negative indicating a possible mixed site occupancy. Therefore, we assumed a slight anti-site disorder on the Ca1 position, and, for reasons of formal charge equilibrium, on the Os2 position (the one Os-position showing a higher thermal displacement parameter). About 20 % of the Ca1 position could be occupied by $Os^{5+}$, and in consequence about 10 % of the Os2 position by $Ca^{2+}$. The structure parameters, atomic positions and important bond lengths obtained from neutron data are listed in Table S1, S3 and S4 respectively, in SI.

Rietveld refinement of $Sr_3CaRu_2O_9$ phase was carried out to check the phase purity; the respective profile plot is shown in Figure S4 in SI.

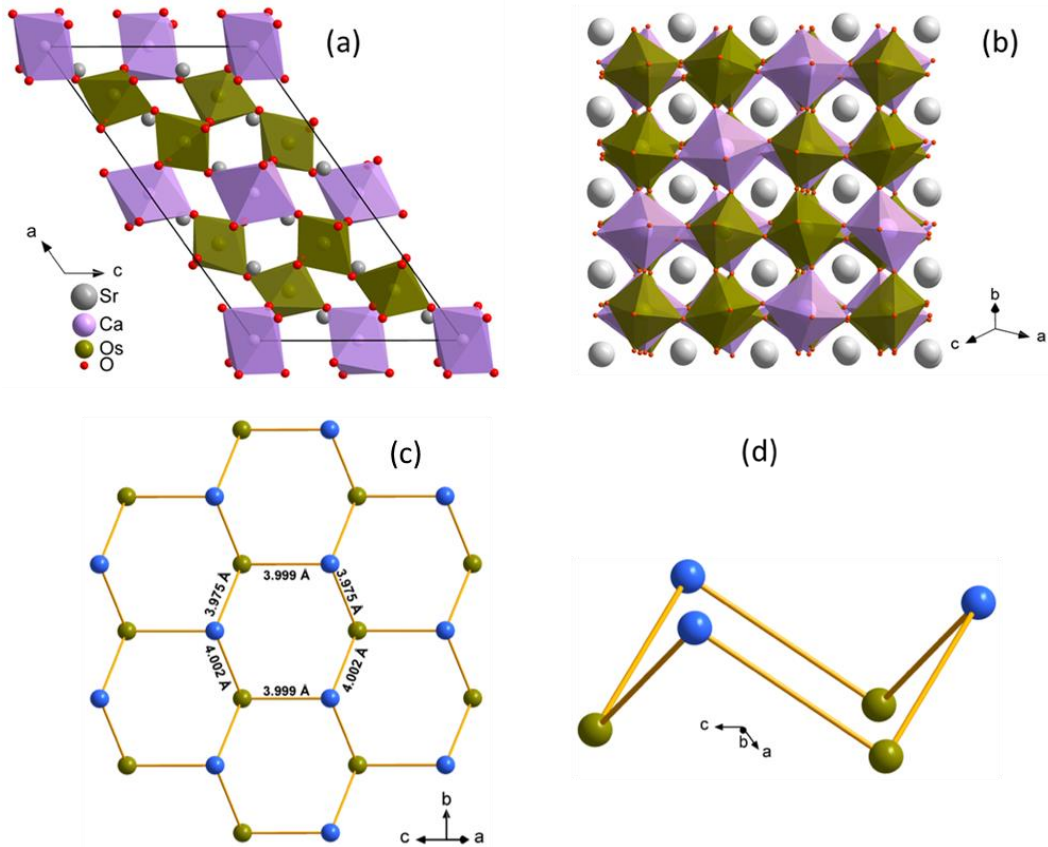

Figure 3: Projections of the crystal structure of $Sr_3CaOs_2O_9$ in different crystallographic directions (a and b), buckled honeycomb lattice of Os (c and d). Os1 (olive) and Os2 (blue) are labeled with different colors in (c and d) only for the purpose of clarity.

Figure 3 shows the crystal structure of $Sr_3CaOs_2O_9$ in different crystallographic orientations. It is isostructural to $Sr_3CaM_2O_9$ (M = Ru and Ir) and is best described as a 2:1 ordered triple perovskite were Ca and Os both occupy distinct crystallographic positions. The $CaO_6$ octahedra are corner shared with $OsO_6$ and occur alternatively after every pair of corner-shared $OsO_6$ octahedra, hence the ordering scheme 2:1. This cation ordering is driven by a large charge difference of 3 and ionic radii mismatch by ~ 0.45 Å between $Ca^{2+}$ and $Os^{5+}$ cations.[35] Because of the distortions driven by deviation from the ideal cubic perovskite structure, significant tilting and distortion of the $CaO_6$ and $OsO_6$ octahedra is observed. The corner shared connectivity pattern of tilted $OsO_6$ octahedra gives rise to a unique 2D buckled honeycomb lattice of Os where the average distance between the two adjacent Os atoms is ~4.0 Å (Figure 3c and d). This kind of corner shared buckled honeycomb lattice has been observed before only in the isostructural compounds $Sr_3Ca(Ru/Ir)_2O_9$.[10,11] Moreover, honeycomb pattern of Os is even rarer and in fact has only been reported previously in $Li_{2.15}Os_{0.85}O_3$ (although, edge-shared).[36] Bond distances between Os and O are in the range of 1.88-2.06 Å and agree well with those reported in literature.[22,37–40] However, the Ca-O distances in $Sr_3CaOs_2O_9$ appear to be slightly shorter (2.19-2.24 Å) than expected, as far as their ionic radii are concerned[35]. These short Ca-O distances comply with a small degree of anti-site disorder at the Ca and Os positions. Such shortened distances also appear in the two isostructural Ir and Ru compounds indicating that there might exist a similar slight B site disorder, which however, is not explicitly mentioned in those reports.[11,12] It is important to note that not many perovskites containing Os are known and most of them contain Ba as an A cation. While most of the Ba containing triple perovskites derive from the aristotype hexagonal perovskite, the Sr containing representatives tend to form variants of the ideal cubic perovskite structure.[11,12]

**3.2 Magnetism, conductivity, specific heat.** Figure 4 displays the temperature dependent susceptibility and isothermal magnetization data of $Sr_3CaOs_2O_9$. The magnetic susceptibility of $Sr_3CaOs_2O_9$ is small and exhibits an anomaly at $T_N = 385$ K (Fig. 4(a)) which we assign to the antiferromagnetic ordering of the Os moments. Below the Néel temperature, $\chi_m(T)$ decreases (inset of Fig. 4(a)) but increases again below $\approx 150$ K, indicating possible paramagnetic impurities (Curie tail). Moreover, the overall field dependence suggests some trace ferromagnetic contamination with high Curie temperature. Above $T_N$, the corrected susceptibility is paramagnetic but does not follow a clear Curie-Weiss law, therefore, it was not possible to determine the paramagnetic effective moment. The linear field dependence of the magnetization at $T = 1.8$ K confirms the antiferromagnetic nature of the ordering. The para- and ferromagnetic contributions are invisible on this scale.

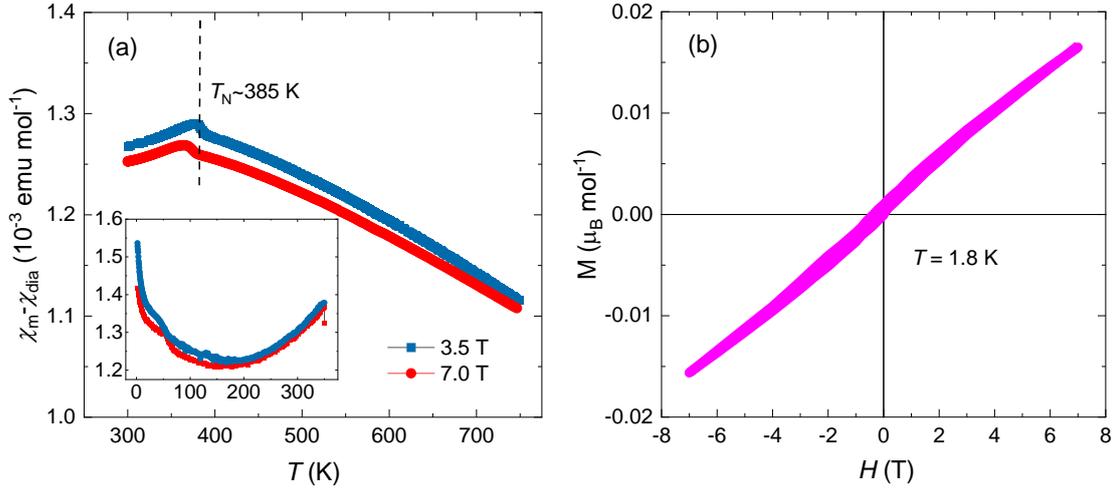

Figure 4: (a) Temperature dependence of molar susceptibility in high temperature range (300-750 K) and (b) isothermal magnetization of $Sr_3CaOs_2O_9$. Inset of (a) shows the molar susceptibility data in the $T$ range of 2-350 K.

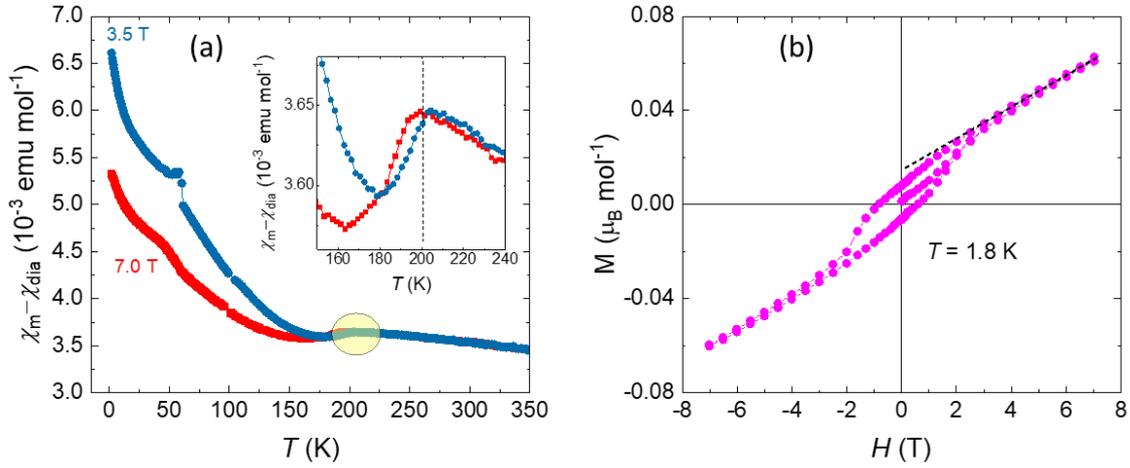

Figure 5: (a) Temperature dependence of molar susceptibility of $Sr_3CaRu_2O_9$ at various fields in FC mode and (b) isothermal magnetization of $Sr_3CaRu_2O_9$ at $T = 1.8$ K. The inset of (a) shows an enlarged view of susceptibility near the magnetic transition $\approx 200$ K.

$Sr_3CaRu_2O_9$ appears to be predominantly antiferromagnetically ordered with a transition temperature of 200 K (Figure 5). There are few tiny ferromagnetic like transitions also at lower temperature, most prominent one at ~170 K. An estimation of the total weak ferromagnetic component was made based on isothermal magnetization studies. A very small value of saturated moment ~0.007 $\mu_B$/Ru$^{5+}$ is obtained which might be contributed by traces of ferromagnetic impurity, most probably $SrRuO_3$ (~0.5 %) as adjudged by the weak transition at ~165 K observed at low fields (not shown). This further supports principle antiferromagnetic ordering of the bulk sample. The susceptibility in paramagnetic state does not follow a Curie-Weiss law, preventing the estimation of any effective moment or the strength of magnetic interactions. Figure 6 shows the specific heat data of $Sr_3CaOs_2O_9$.

The magnetic ordering in $Sr_3CaOs_2O_9$ at ~385 K is confirmed by heat capacity measurement, which clearly shows a pronounced λ-shaped anomaly centered around 385 K. A jump of ~10 J/mol•K is observed. An antiferromagnetic ordering at such high temperature is rare in perovskite compounds. Notably, the other isostructural compound $Sr_3CaIr_2O_9$ displays paramagnetic behavior with a vanishingly small susceptibility associated with a possible $J_{eff} = 0$ ground state of Ir$^{+5}$ ions,[12] whereas $Sr_3CaRu_2O_9$ is found to be antiferromagnetic below 200 K. A growing body of high temperature magnetic compounds containing osmium opens new avenues to explore high temperature magnetism in 5$d$ compounds and especially in otherwise scarcely investigated osmates.

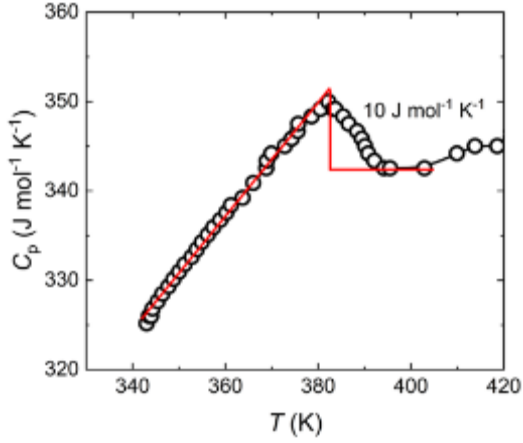

Figure 6: Specific heat near the expected transition of $Sr_3CaOs_2O_9$ showing a pronounced λ-shaped anomaly at the magnetic transition temperature.

Figure 7 shows the DC electrical resistivity of $Sr_3CaOs_2O_9$ in the temperature range 115 to 300 K. Room temperature resistivity, $\rho^{300\,K} \sim 1$ kΩ•cm indicates the material is electrically insulating in nature and exhibits a activated-type behavior at low temperatures with resistivity increasing by at least three orders of magnitude. A simple Arrhenius fit in the high temperature region was attempted to extract a band gap ($E_g = 0.35$ eV), however the curve is linear only in a small range of high temperature (225 K – 300 K) and deviates from linearity at lower temperatures. Rather the resistivity fits almost perfectly to $\rho(T) \propto [\exp(\Delta/T)]^{1/4}$ equation in almost entire temperature range (figure 7 inset). This is indicative of a three-dimensional hopping of charge carriers.

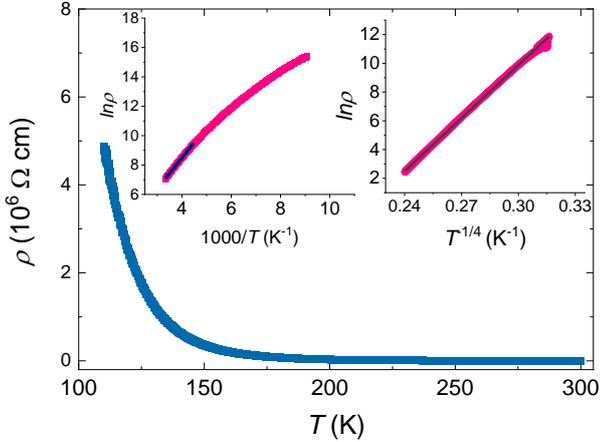

Figure 7: Resistivity vs temperature curves for $Sr_3CaOs_2O_9$. Left and right insets show the $\ln\rho$ vs $1/T$ and $\ln\rho$ vs $1/T^{1/4}$ plots respectively, along with the linear fits (black line).

**3.3 Neutron Diffraction and magnetic structure.** To determine the magnetic structure, we collected neutron diffraction data above and below the expected magnetic transition temperature. The data below $T_N$ shows at least two nuclear Bragg peaks with additional intensity which can be assigned to magnetic ordering. The curve in Figure 8 shows the variation of intensity of the peak, containing the overlapping Bragg reflections (-111) and (-202), with temperature. This also confirms the $T_N$ of ~385 K observed in the susceptibility and heat capacity studies. The magnetic structure however could not be solved unambiguously, resulting in two models with almost equally good fits as explained below.

All magnetic scattering intensity lays on top of nuclear peaks, and the propagation vector is κ = (000). This propagation vector, together with the known crystal symmetry of the chemical unit cell and the two positions of the $Os^{5+}$ magnetic ions serve as input for the symmetry analysis of possible magnetic ordering schemes. The representation analysis of the low-temperature magnetic structure was performed with the program BasIreps[41–43] as part of the FullProf suite.[44] All predicted symmetry allowed spin configurations were checked on the observed data with Rietveld refinements by FullProf.[45,46] A difference pattern from data acquisitions of two hours each at 2.41 Å at 1.8 and 395 K, respectively (figure 9), has been used at this purpose, where the high temperature data have been corrected for thermal expansion and re-interpolated accordingly. Only four out of 4•2 = 16 combinations of the four irreducible representations (see table S5) for each one of the two $Os^{5+}$ ions provide sets of calculated diffraction peaks consistent with our experimental data. Two of them refine to purely ferromagnetic structures with all spins parallel and must be excluded according to the susceptibility data. The remaining two representations, however, cannot be distinguished unambiguously by Rietveld refinement, despite a small preference for one of them.

A few assumptions need to be made hereafter, due to the low number of observations (mainly two unambiguous Bragg peak intensities) and the possible number of parameters describing the magnetic ordering. First, we assume that the $Os^{5+}$ ions on both positions show the same magnetic moment. The reason why the two hereafter described solutions cannot be unambiguously distinguished is the accidental overlap of the (-111) and (-202) reflections. The two reflections are only 0.04° apart from each other in 2θ at a wavelength of 2.4 Å, which is virtually indistinguishable by any neutron diffractometer, it would require a resolution of Δ$d$/$d$ of about $10^{-3}$, which is technically inaccessible today, by an order of a magnitude. The only other distinctive feature is the intensity ratio of the bespoken (-111) and (-202) peak to a less intense, higher angle peak regrouping (220), (-513) and (113).

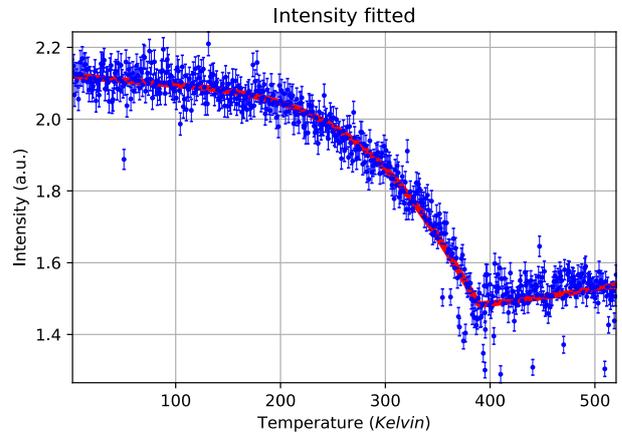

Figure 8: Temperature dependence of intensity of the magnetic and nuclear peak of $Sr_3CaOs_2O_9$ containing the overlapping Bragg reflections (-111) and (-202).

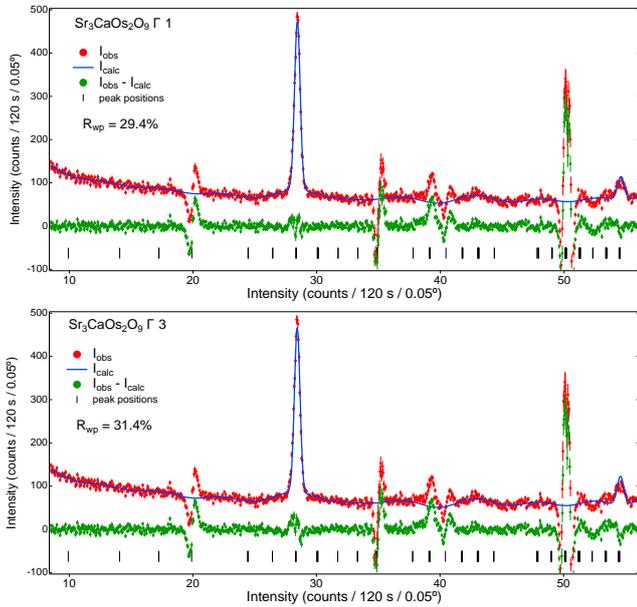

Figure 9: Rietveld fit (blue line) (difference of observed minus calculated intensity, green dots) of the difference pattern (pattern at 1.8 K minus isotropic thermal expansion-corrected pattern at 395 K (red dots with estimated error bars) with the pure magnetic phase. Positions of strong nuclear peaks are excluded (light blue background) because of a remaining misfit due to thermal expansion. Black lines indicate Bragg peak positions.

One solution, with a more intense (-111) peak, corresponds to the representation Γ 1 (Shubnikov group P2$_1$/c', #14.78) for both Os$^{5+}$ ion positions with main spin vector components perpendicular to [010] and tilted to the plane of buckled honeycomb lattice of osmium. Another one, with a more intense (-202) peak, corresponds to Γ 3 (Shubnikov group P2$_1$'/c, #14.77), with main vector components parallel or antiparallel to [010] and, thus, inside the plane of the puckered honeycomb lattice of osmium. The two solutions shall be called "Γ 1" and "Γ 3" hereafter. Both magnetic structures are best described using spherical coordinates, as this allows for Rietveld refinement with a single constrained magnetic moment for both osmium positions, thus, with five independent parameters describing the two vectors, rather than six.

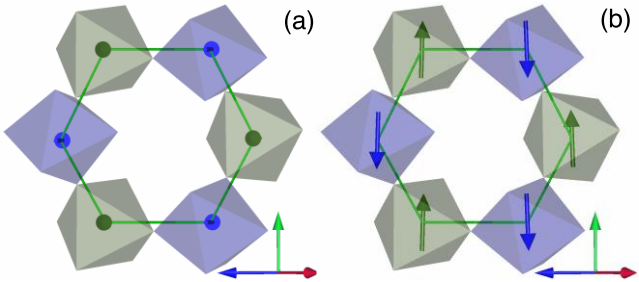

Figure 10: View perpendicular to the puckered planes: magnetic moments of osmium in hexagonal neighborhood, Os1 oxygen-coordination polyhedra olive, Os2 blue, oxygen positions form octahedra corners. (a) First solution (Γ 1) with vectors out of the puckered lattice plane, (b) second solution (Γ 3) with vectors parallel to b-axis.

In the solution Γ 1, the magnetic moments are tilted to the puckered honeycomb lattice plane, defined by the positions of either osmium (Figure 10a and 11a). The vectors are tilted by 66° for the moments of the Os1 positions and by 64° for Os2. Whereas the moments are not fully perpendicular to this plane, they are almost perpendicular to the direction of the b-axis. They are tilted to the ac-plane normal to the b-axis by 5.6° for moments of Os1 and by 4.2° for those of Os2. In the buckled planes, each osmium has three closest neighbors of the other osmium position, and six osmium positions form a puckered (chair conformation) hexagonal ring, alternatingly by Os1 and Os2. Os1 positions are on one side of the plane, Os2 on the other. In a plane, all osmium moments belonging to one position point in roughly the same direction with a deviation from perfect parallel alignment by 8.8° for Os1 and 8.4° for Os2, respectively. The angle between next neighbors is 171.0° for the two neighbors of each Os-position in direction of the b-axis (or the four out of six edges of the buckled hexagonal ring not perpendicular to the b-axis) and 177.5° for the remaining third of closest Os-Os distances, i.e., those 'bonds' perpendicular to the direction of the b-axis.

In the solution Γ 3 (figure 10b and 11b), the magnetic moments are lying in the puckered honeycomb lattice plane, defined by the positions of either osmium atoms. The vectors are tilted by only 0.6° for the moments of the Os1 positions and by 0.1° for Os2. They derive from the direction of the b-axis by 5.8° for moments of Os1, and 4.7° for those of Os2, respectively. In a puckered honeycomb lattice plane, all osmium moments belonging to one position point in roughly the same direction with a deviation from perfect parallel alignment by 11.6° for Os1 and 9.5° for Os2. The angle between next neighbors is 169.5° for the two neighbors of each Os-position in direction of the b-axis (or the four out of six edges of the puckered hexagonal ring not perpendicular to the b-axis) and 178.8° for the remaining third of closest Os-Os distances, i.e., those 'bonds' perpendicular to the direction of the b-axis.

Regarding the large error bars on the angles of both solutions which exceed largely the aforementioned deviations from collinearity or perpendicularity we can consider further assumptions to obtain a less correlated fitting result.

The only angle that needs refinement is the moment orientation perpendicular to the ac-plane in the solution Γ 1. All other angles mentioned should be constrained to be 0°, 90° or 180° respectively. In the case of Γ 1 this results in refining only the x and z component of the moments, constrain them to the same amount over the two positions and set the y component zero, respectively, in spherical coordinates, fixing the azimuthal φ angle to zero and constrain the amount of the inclination angle θ to be the same for both positions, thus, two rather than five independent parameters. Even though, the error bar on the remaining inclination angle in the spherical representation remains enormous. The magnetic moment vectors are standing at a 71.6° angle on the buckled planes. The magnetic moment amounts here to 1.41(9) μ$_B$. The inclination angle θ can further be constrained to be 90° - β, the monoclinic angle without degradation of the fit, and resulting in a magnetic moment of 1.419(8) μ$_B$. The moment vectors then form an angle of 70.5° with the buckled planes.

In the solution Γ 3 no angle remains to be refined, and in the Cartesian description as well only one parameter, the amount of magnetic moment remains to be refined, the *y*-component of the Cartesian vector. It must be noted here that the quality of refinement is equal to the one with five or six refinable parameters. The magnetic moment amounts here to 1.88(1) $\mu_B$. See table S6 for the two solutions and S7 for more details.

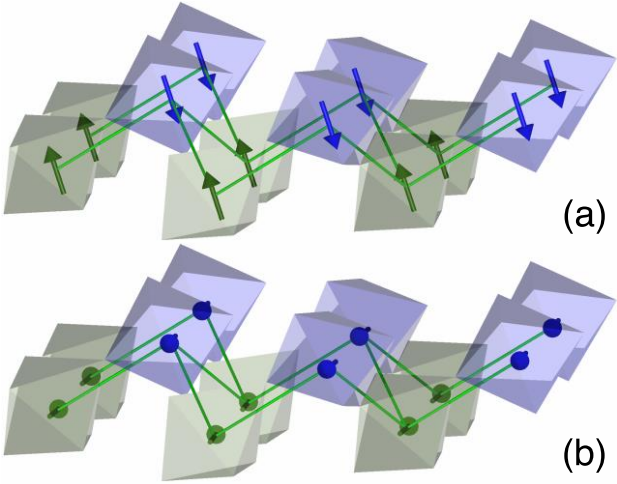

Figure 11: View in direction of the *b*-axis: magnetic moments of osmium in hexagonal neighborhood, Os1 coordination octahedra olive, Os2 blue, oxygen positions form octahedra corners. (a) First solution (Γ 1) with vectors tilted to the puckered planes, (b) second solution (Γ 3) with vectors parallel to *b*-axis.

**3.4 DFT analysis.** Despite the thorough experimental characterization of $Sr_3CaOs_2O_9$, two questions remain open: what causes its unusually high long-range antiferromagnetic ordering temperature and how the Os magnetic moments are oriented with respect to the crystal axes.

We start with the latter problem. Our refinements of neutron diffraction data reveal two possible solutions: i) Γ 1 with Os moments oriented nearly perpendicular to the puckered planes (Fig. S5a) or ii) Γ 3 with Os moments lying in-plane, parallel to the *b* crystal axis. To distinguish between these two solutions, we performed full-relativistic total energy GGA+$U$ calculations for the respective magnetic configurations. The Γ 1 configuration has the lowest energy, separated from the Γ 3 configuration by 2.32 meV/Os (Table S8). Band gaps of Γ 1 and Γ 3 amount to 1.097 (1.062) eV. The lengths of the magnetic moment of Os amount to 1.842 (1.780) $\mu_B$ for Γ 1 and Γ 3, respectively.

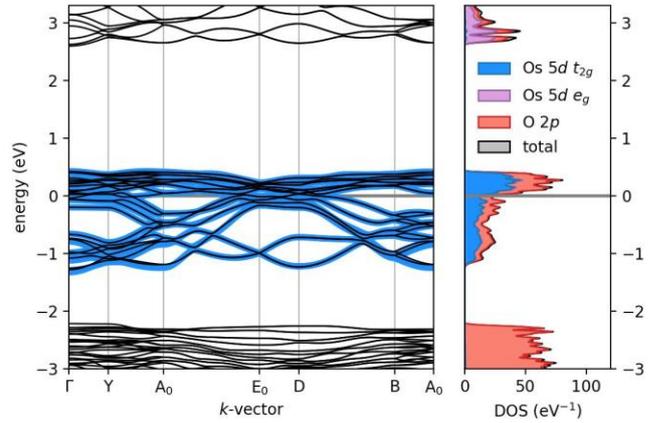

Figure 12: (Left) The nonmagnetic GGA band structure of $Sr_3CaOs_2O_9$ (thin black lines) and the Fourier-transform of the Wannier Hamiltonian (thin blue lines) obtained by projecting the GGA bands onto Os 5*d* $t_{2g}$ orbitals. (Right): Density of states (DOS) for the same calculation. O 2*p*, Os 5*d* $t_{2g}$ and $e_g$ contributions are highlighted.

To understand the root cause of the high magnetic ordering temperature, we perform a nonmagnetic DFT calculations. The band structure (Fig. 12, left panel) features a 24-bands manifold which has a dominantly Os 5*d* $t_{2g}$ character with an admixture of O 2*p* states. This manifold is well separated by sizable band gaps from the rest of the valence band dominated by O 2*p* states as well as the bottom of the polarization band having the dominant Os $e_g$ contribution (Fig. 12, right panel). The nonzero density of states of the Fermi level indicates the metallic behaviour, at odds with the experiment. This discrepancy stems from the underestimation of electronic correlations and the neglect of the spin-orbit coupling in the Os 5*d* shell. The insulating state is readily restored in our full-relativistic DFT+$U$ calculations that are discussed below.

Despite the spurious metallic ground state, the nonmagnetic band structure of $Sr_3CaOs_2O_9$ harbours information on the electron transfer integrals that underlie the magnetic exchange. To extract this information, we project the 24-bands manifold onto a Wannier basis of the $t_{2g}$ orbitals of Os. In this way, we obtain the transfer integrals between the pertinent Os *5d* orbitals (*xy*, *yz*, *xz*) for each pair of Os atoms directly from the respective matrix elements. First, we consider the transfer integrals within the puckered honeycomb planes. We find that each of three nearest-neighbour Os-Os bonds features two sizable transfer integrals ranging between 220 and 300 meV (Table S9). These pertain to $t_{2g}$ orbitals of Os having the same orbital character and facilitated by O 2*p* contributions as can be seen in Fig. 13 (left and central panels). Hence, a sizable antiferromagnetic exchange between the nearest neighbours can be expected. The third pair of orbitals is essentially decoupled, as evidenced by the minute O 2*p* contribution of the respective Wannier function (Fig. 12, right panel). Transfer integrals between second neighbours amounts to 90-110 meV, all longer-range terms within the buckled planes are considerably smaller. The presence of second-neighbour exchange indicates that the buckled layers are magnetically frustrated.

Interlayer hoppings of 120-130 meV are facilitated by two inequivalent paths with Os-Os distances of 5.789 and 5.818 Å (Fig. S6). Hence, we can expect a sizable and non-frustrated interlayer exchange. To estimate this exchange quantitatively, we turn to full-relativistic DFT+$U$ calculations of different magnetic configurations. We construct an antiferromagnetic configuration with flipped Os spin orientation of the second plane compared to experimental proposed magnetic structure (AFM-exp) (Fig. S5b); in configuration AFM-exp (AFM-flip) the Os moments in the neighbouring planes are antiparallel (parallel). Then we perform total energy DFT+$U$ calculations for two antiferromagnetic configurations along [101] and [001] directions. The AFM-exp configurations are energetically favourable by 8.83 and 8.89 meV/Os, respectively. We can therefore conclude that $Sr_3CaOs_2O_9$ features a sizable antiferromagnetic exchange between the puckered planes. This corroborates the high long-range antiferromagnetic ordering temperature of this compound.

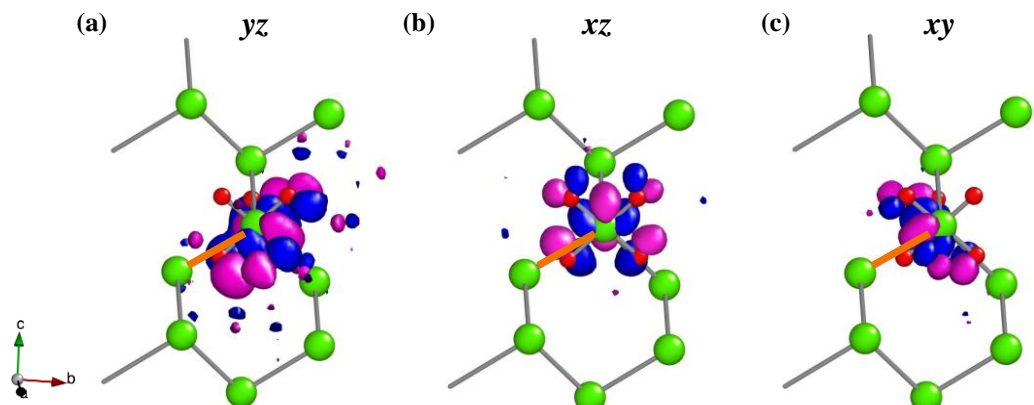

Figure 13: Os-centered Wannier functions with the $t_{2g}$ symmetry. Note the sizable contributions of O 2$p$ orbitals. For the highlighted (orange line) pair of Os atoms, sizable hopping is facilitated only by the $yz$ and $zx$ orbitals. Green and red spheres are Os and O, respectively.

## 4. CONCLUSIONS

In conclusion we demonstrate here structural and magnetic properties of a new triple perovskite $Sr_3CaOs_2O_9$. It shows a 2:1 ordering at the B sites and is thus the first example of osmate perovskites exhibiting such ordering. $Sr_3CaOs_2O_9$ shows long range antiferromagnetic ordering at an unexpectedly high $T_N$ of ~385 K, which is confirmed by calorimetric and neutron diffraction studies. Such high $T_N$ is surprising in a compound with only 5$d$ magnetic atoms and is attributed here to the sizeable interlayer interaction. Thus, the recently growing body of Os based perovskites with high transition temperatures offers exciting avenues for discovering interesting magnetic materials and studying their intriguing interplay of spin, orbital and lattice degrees of freedom.

## ASSOCIATED CONTENT

Supporting Information.
Elemental analysis, thermal analysis, additional figures, and crystallographic tables are presented in supplementary information. "This material is available free of charge via the Internet at http://pubs.acs.org."

## AUTHOR INFORMATION

### Corresponding Author

* M.Jansen@fkf.mpg.de

### Author Contributions

The manuscript was written through contributions of all the authors. All authors have given approval to the final version of the manuscript.

### Notes

The authors declare no competing financial interest.

### ACKNOWLEDGMENT

GST thank the Würzburg-Dresden Cluster of Excellence *ct.qmat* (EXC 2147) funded by the *Deutsche Forschungsgemeinschaft* (DFG) for partial financial support. The authors acknowledge financial support from the German Research Foundation (Deutsche Forschungsgemeinschaft, DFG) via SFB1143 Project No. A5 and under Germany's Excellence Strategy through Würzburg-Dresden Cluster of Excellence on Complexity and Topology in Quantum Matter—ct.qmat (EXC 2147, Project No. 390858490). The authors thank Marcus Schmidt for TGA and DSC measurements. The authors thank Ulrike Nitzsche for technical assistance.